\documentclass{article}
\usepackage{amssymb}

\usepackage{graphicx}
\usepackage{amsmath}

\input{tcilatex}

\begin{document}

\title{Dynamic critical exponents of the Ising model with multispin interactions}
\author{C. S. Sim\~{o}es$^{1}$ and J. R. Drugowich de Fel\'{i}cio$^{1,2}$ \\
1-Departamento de F\'{i}sica e Matem\'{a}tica - FFCLRP -\\
Universidade de S\~{a}o Paulo\\
Av. dos Bandeirantes, 3900 - Ribeir\~{a}o Preto, S\~{a}o Paulo, Brasil\\
2-Faculdades COC - \\
Rua Abrah\~{a}o Issa Hallack, 980-\\
Ribeir\~{a}o Preto-, S\~{a}o Paulo, Brasil}
\maketitle

\begin{abstract}
We revisit the short-time dynamics of 2D Ising model with three spin
interactions in one direction and estimate the critical exponents $z,$ $%
\theta ,$ $\beta $ and $\nu $. Taking properly into account the symmetry of
the Hamiltonian we obtain results completely different from those obtained
by Wang et al.. For the dynamic exponent $z$ our result coincides with that
of the 4-state Potts model in two dimensions. In addition, results for the
static exponents $\nu $ and $\beta $ agree with previous estimates obtained
from finite size scaling combined with conformal invariance. Finally, for
the new dynamic exponent $\theta $ we find a negative and close to zero
value, a result also expected for the 4-state Potts model according to Okano 
\textit{et al}..
\end{abstract}

\qquad \qquad \qquad \qquad \qquad

{\small \bigskip }

Since the work by Janssen \textit{et al }\cite{Janssen} and Huse \cite{Huse}
pointing out the existence of another universal stage at an early time
critical dynamic, several statistical models have been investigated to
confirm the analytical predictions about the ''critical initial slip'' and
to enlarge the knowledge of critical phenomena \cite{Li1995}, \cite{Li}, 
\cite{Okano}, \cite{Luo}, \cite{Schulke}, \cite{Simoes}. The investigation
of the universal behavior in short-time dynamics avoids the critical slowing
down effects of the equilibrium and provides an alternate way \cite{Zheng}
to calculate the new exponent $\theta $ which governs the behavior of the
magnetization, the dynamic critical exponent $z$ as well as the static
exponents $\beta $ and $\nu $.

In this letter we adopt this approach to study the two-dimensional (2D)
Ising model with three-spin interactions in one direction and calculate its
set of exponents. Motivation came from the fact that in a recent paper by
Wang \textit{et al }\cite{Wang} estimates obtained were in complete
disagreement with pertinent results. Here we show that, when the symmetry of
the model is taken properly into account, good agreement is obtained with
expected results.

The Hamiltonian of the 2D Ising model with three spin interactions ($m=3$)
in one direction is \cite{Turban1982} \ \ \ \ \ \ \ \ \ \ \ \ \ \ \ \ \ 

\begin{equation}
-\beta H=\sum\limits_{<i,j>}\left\{
K_{y}S_{i,j}S_{i+1,j}+K_{x}\prod_{l=0}^{m-1}S_{i,j+l}\right\}
\label{multispin}
\end{equation}
where $S_{i,j}=\pm 1$ is the Ising spin variable. The model is known to be
self-dual \cite{Debierre}, its critical line being

\begin{equation}
\sinh 2K_{x}\sinh 2K_{y}=1  \label{linha_critica}
\end{equation}
for all $m$. For the particular isotropic case ($K_{x}=K_{y}$) the critical
coupling is 
\begin{equation}
K_{c}=J/k_{B}T_{c}=\frac{1}{2}\ln (\sqrt{2}+1)=0.4406867,
\label{temp_critica}
\end{equation}
which is the same of the standard 2D Ising model.

Symmetry analysis and ground state degeneracy considerations suggest that
the model is in the same universality class as the $q$-state Potts model,
whenever $q=2^{m-1}$ \cite{Debierre}, \cite{Caticha}. This result is
supported by finite-size scaling studies \cite{Penson}, \cite{Debierre}, 
\cite{Vanderzande}, weak and strong coupling expansions \cite{Igloi},
conformal invariance \cite{Kolb}, \cite{Igloi1987}, \cite{Penson1988},
standard Monte Carlo simulations \cite{Blote}, \cite{Alcaraz1987}\cite
{Caticha} and mapping of the $m=3$ model in the extreme anisotropic limit of
the $4$-state Potts model \cite{Blote1987}. In most of those papers the
argument to include the Ising model with three-spin interaction and the
4-state Potts model in the same universality class is based on the value of
the exponents $\nu $ and $\alpha $($\approx 2/3$) \cite{Zhang}. All of them
respect the symmetry of the Hamiltonian. Very little is known about the
exponents $\beta $ \cite{Alcaraz1987} and $z$ \cite{Wang}.

The ground state for the general $m$-spin interaction is $2^{m-1}$%
degenerated \cite{Penson}, which implies that the 2D Ising model with $m=3$
spin interaction is four-fold degenerated. The relevant symmetry of this
model is semi-global \cite{Alcaraz} and the Hamiltonian is symmetric under
the reversal of all the spins in any two sublattices, which leads to the
existence of three independent interpenetrating sublattices in the system.
At $T=0$, the possible states consist of repetitions of the patterns $+++,$ $%
+--,$ $-$ $+-,$ $--+$ in the horizontal direction, copied along the lines.
According to preceding statement, it is important to take lattice sizes that
are multiple of three - in order to respect the symmetry of the Hamiltonian
- and to work with the appropriate order parameter - the sublattice
magnetization - in order to avoid the effect of staggered magnetization. We
suspected that Wang \textit{et al} \cite{Wang} didn't consider the $m$-spin
symmetry in their simulations\ neither worked with the magnetization of
sublattice, since they just presented results for square lattices which do
not obey that previous condition (multiple of three). In this sense, it
seemed to us relevant to repeat simulations, paying attention to the above
mentioned points, to check the apparent failure of the short-time approach
in this case.

We began by repeating the analysis made by Wang \textit{et al} \cite{Wang}
for the Binder cummulant

\begin{equation}
U(t,L)=\frac{\left\langle M^{2}(t)\right\rangle }{\left\langle
M(t)\right\rangle ^{2}}-1
\end{equation}
where $<$ $>$ means average on samples, $\ t$ is the time, $%
M(t)=\sum\limits_{i=1}^{L}\sum\limits_{j=1}^{L}S_{i,j}$ is the magnetization
at time $t$ and $L$ the size of the square lattice. They argue that this
expression obeys the power law form:

\begin{equation}
U(t,L)\varpropto t^{d/z}\text{,}  \label{d/z}
\end{equation}
when the dynamical process starts from an ordered state ($m_{0}=1$), which
is a fixed point under renormalization group transformation. Fig. 1 shows
explicitly the different results obtained when we use the magnetization of
the sublattice. The average is taken over 50000 independent initial
configurations and the error bars (smaller than the size of the points) are
obtained by repeating five times each simulation. When the simulation is
performed without the sublattice considerations the slope of the curve
agrees with that presented by Wang \textit{et al }and confirmed our
suspicions. From our point of view \cite{Drugo et al}, though, this
cummulant should obey the power law form (\ref{d/z}) only when different
initial conditions were used in the study of the magnetization and its
second moment. Scaling arguments \cite{Li1995} assert that the second moment
of magnetization behaves as 
\begin{equation}
\left\langle M^{2}(t)\right\rangle \varpropto t^{(d-2\beta /\nu )/z}\text{,}
\label{2nd_moment}
\end{equation}
only when the samples are taken with zero initial magnetization ($m_{0}=0)$.
On the other hand, short-time scaling behavior implies 
\begin{equation}
\left\langle M(t)\right\rangle \varpropto t^{-\beta /\nu z}  \label{mag}
\end{equation}
for samples starting from the ordered state ($m_{0}=1$) \cite{Li1996}. Thus,
performing two different simulations under those mentioned conditions, we
obtain the time evolution of the ratio $\left\langle M^{2}(t)\right\rangle
/\left\langle M(t)\right\rangle ^{2}$ which furnishes the exponent $d/z$ in
a log-log plot (Fig. 2). From the slope of that curve we estimate $%
z=2.380\pm 0.004$. \qquad \qquad

To confirm our result we used two other approaches: the generalized fourth
order Binder's cumulant \cite{Li1995} and the parameters Q and R introduced
by de Oliveira\cite{Murilinho}. In both cases, the collapse of \ the curves
for different lattice sizes, at critical temperature, are used to determine
the dynamical critical exponent $z$ from short-time simulations. The
Binder's cummulant, 
\begin{equation}
U_{4}(t,L)=1-\frac{\left\langle M^{(4)}\right\rangle }{3\left\langle
M^{(2)}\right\rangle ^{2}}
\end{equation}
satisfies, at $T=T_{c}$ the scaling relation

\begin{equation}
U_{4}(t,L)=U_{4}(b^{-z}t,b^{-1}L)
\end{equation}
where $b=L/L^{\prime }$ since $U_{4}$ scales as $L^{0}$. This technique has
proved to be useful in determining the exponent $z$ and was applied to the
2D and 3D Ising models \cite{Li1995}, \cite{Schulke e Zheng 3D}, the 3-state
Potts model \cite{Schulke}, the majority vote model \cite{Mendes} and
cellular automata \cite{Drugo e Tania}. The initial magnetization of
samples, in this case, is zero as well as the correlation length. The
results are good enough. Error bars, however, are bigger than those obtained
by damage spreading technique \cite{Grassberger}. In Fig. 3, we show the
collapse of the cumulant $U_{4}$ for lattice pairs ($L,2L$) with $z=2.3$. In
fact, the range of $z$ \ for which the collapse is still observed is given
by $z=2.3\pm 0.1$. This result supports our previous estimate for $z$ ($%
2.383\pm 0.004$) and can be related to the 4-state Potts model exponent \cite
{Arcangelis}. In order to stress the importance of considering the symmetry
of Hamiltonian, we exhibit in Fig. 4 the deformation of the Binder cumulant
when there is no sharp preparation of the initial magnetization on the
sublattices and the magnetization evolution is calculated without
restrictions.

When we use the parameters $Q(t,L)=\left\langle sign\left( \frac{1}{L^{2}}%
M(t)\right) \right\rangle $ \ and $R(t,L)=\left\langle \left( sign\frac{1}{L}%
\sum\limits_{bottom}S_{i}\right) \left( sign\frac{1}{L}\sum%
\limits_{top}S_{i}\right) \right\rangle $ of de Oliveira\cite{Murilinho} and
scaling relations for $T=T_{c}$ \cite{Cesar Barreto}

\begin{equation*}
Q(t,L)=Q(b^{-z}t,b^{-1}L)
\end{equation*}
and\bigskip

\begin{equation*}
R(t,L)=R(b^{-z}t,b^{-1}L)
\end{equation*}
we obtain the collapse among curves (see Fig. 5 and Fig. 6) for different
lattices when time is scaled with $z=2.3$. Samples in these cases were
initialized with all spins up ($m_{0}=1$).

\bigskip In order to calculate the exponent $\nu $, we studied the
derivative of the magnetization, which presents the following scaling form:

\begin{equation}
\partial _{\tau }\ln M(t,\tau )|_{\tau =0}=t^{1/\nu z}\partial _{\tau
^{\prime }}\ln F(\tau ^{\prime })|_{\tau ^{\prime }=0}.
\end{equation}
Fig. 7 shows the power law behavior of the $\partial _{\tau }\ln M(t,\tau )$
when $\Delta \tau =0.0002.$ The measured slope of the curve gives $1/\nu
z=0.624\pm 0.005$. Thus, taking $z=2.383\pm 0.004$ \ we find $\nu =0.67\pm
0.01$,\ to be compared with the conjectured value $\nu =2/3.$

Since the values of the exponents $\nu $ and $z$ are already known, the
exponent $\beta $ of the magnetization can be obtained from the power law
increase of the second moment of magnetization, Eq.(\ref{2nd_moment}). Fig.
8 presents, in a double-log scale, the polynomial behavior of $\left\langle
M^{(2)}(t)\right\rangle $. From the slope of these lines we estimate $\beta
=0.11\pm 0.02$. This result is in agreement with the expected value $\beta
=0.125$ \cite{Alcaraz}, \cite{Ubiraci}, \cite{Turban2}.

In order to extend the picture of universality we investigate the exponent $%
\theta $ by a recent technique proposed by Tom\'{e} and de Oliveira \cite
{Tania}. In their paper they show that the exponent $\theta $ can also be
independently calculated in despite of the sharp preparation of the samples,
since the time correlation of the total magnetization in samples with a
random initial configuration also exhibits polynomial behavior 
\begin{equation}
\left\langle M(t)M(0)\right\rangle =\frac{1}{N}\left\langle
\sum_{i}\sum_{j}S_{i}(t)S_{j}(0)\right\rangle \varpropto t^{\theta }.
\label{correl_mag}
\end{equation}
This procedure avoids the use of an initial state with a nonzero (but small)
magnetization as well as the numerical extrapolation $m_{0}\rightarrow 0$ to
calculate the dynamic exponent $\theta $. Fig. 9 shows, in a log-log scale,
the $\left\langle M(t)M(0)\right\rangle $ behavior for different lattice
sizes. The slope of those curves give us $\theta =-0.03\pm 0.01$ which is
compatible with the conjecture by Okano \textit{et al} for the 4-state Potts
model \cite{Okano}.

In summary, we have obtained static and dynamic critical exponents for the
Ising model with multispin interactions using short-time Monte Carlo
simulations. Our results show that this model and the 4-state Potts one
share the same set of critical exponents even at dynamic level. When
compared to the paper by Whang \textit{et al} this letter shows the
importance of taking properly into account the symmetry of the Hamiltonian
to deal with magnetization and boundary conditions. We stress that the
present result for the exponent $\nu $ is better than previous estimates
obtained by finite size scaling and Monte Carlo approach \cite{Alcaraz}.


\bigskip

\begin{thebibliography}{99}
\bibitem{Janssen}  H. K. Janssen, B. Shaub and B. Schmittmann, \textit{Z.
Phys.} \textbf{B73,} 539 (1989)

\bibitem{Huse}  D. Huse, Phys. Rev. \textbf{B40}, 304 (1989)

\bibitem{Li1995}  Z. B. Li, L. Sch\"{u}lke and B. Zheng, Phys. Rev. Lett. 
\textbf{74,} 3396 (1995)

\bibitem{Li}  Z. B. Li, X. W. Liu, L. Sch\"{u}lke and B. Zheng, Physica 
\textbf{A245,} 485 (1997)

\bibitem{Okano}  K. Okano, L. Sch\"{u}lke, K. Yamagishi and B. Zheng, J.
Phys. A: Math. Gen. \textbf{30,} 4527 (1997)

\bibitem{Luo}  H. J. Luo, L. Shulke and B. Zheng, Phys. Rev. Lett. \textbf{%
81,} 180 (1998)

\bibitem{Schulke}  L. Sch\"{u}lke and B. Zheng, Phys. Lett. \textbf{A204,}
295 (1995)

\bibitem{Simoes}  C. S. Simoes and J. R. Drugowich de Fel\'{i}cio, J. Phys.
A: Math. Gen. \textbf{31,} 7265 (1998)

\bibitem{Zheng}  B. Zheng, Int. J. Mod. Phys. \textbf{B12,} 1419 (1998)

\bibitem{Wang}  L. Wang, J. B. Zhang, H. P. Ying and D. R. Ji, Mod.\ Phys.
Lett. \textbf{B13,} 1011 (1999)

\bibitem{Turban1982}  L. Turban, J. Phys. C: Solid State Phys. \textbf{15,}
L65 (1982)

\bibitem{Debierre}  J. M. Debierre and L. Turban, J. Phys. A: Math. Gen. 
\textbf{16,} 3571 (1983)

\bibitem{Caticha}  N. Caticha, J. Chahine and J. R. Drugowich de
Fel\'{i}cio, Phys. Rev. \textbf{B43,} 1173 (1991)

\bibitem{Penson}  K. A. Penson, R. Jullien and P. Pfeuty, Phys. Rev. \textbf{%
B26,} 6334 \textbf{\ }(1982)

\bibitem{Vanderzande}  C. Vanderzande and F. Igl\'{o}i, J. Phys. A: Math.
Gen. \textbf{20,} 4539 (1987)

\bibitem{Igloi}  F. Igloi, D. V. Kapor, M. Skrinjar and J. Solyom, J. Phys.
A: Math. Gen. \textbf{19, }1189 (1986)

\bibitem{Kolb}  M. Kolb and K. A. Penson, J. Phys. A: Math. Gen. \textbf{19,}
L779 (1986)

\bibitem{Igloi1987}  F. Igl\'{o}i, J. Phys. A: Math. Gen. \textbf{20,} 5319
(1987)

\bibitem{Penson1988}  A. K. Penson, J. M. Debierre and L. Turban, Phys. Rev. 
\textbf{B37,} 7884 (1988)

\bibitem{Blote}  H. W. Bl\"{o}te, A. Compagner, P. A. M. Cornelissen, A.
Hoogland, F. Mallezie and C. Venderzande, Physica \textbf{A139,} 395 (1986)

\bibitem{Alcaraz1987}  F. C. Alcaraz, Phys. Rev. \textbf{B34,} 4885 (1987)

\bibitem{Blote1987}  H. W. Bl\"{o}te, J. Phys. A: Math. Gen. \textbf{20,}
L35 (1987)

\bibitem{Zhang}  G. M. Zhang and C. Z. Yang, Phys. Rev. \textbf{B48}, 9847
(1993)

\bibitem{Alcaraz}  F. C. Alcaraz and M. N. Barber, J. Phys. A: Math. Gen. 
\textbf{20}, 179 (1987)

\bibitem{Drugo et al}  Roberto da Silva, C. Sim\~{o}es, E. Arashiro, J. R.\
Drugowich de Fel\'{i}cio and N. A. Alves, to be published.

\bibitem{Li1996}  Z. B. Li and L. Sch\"{u}lke, Phys. Rev. \textbf{E53,} 2940
(1996)

\bibitem{Murilinho}  P. M. C. de Oliveira, Europhys. Lett. \textbf{20}, 621
(1992)

\bibitem{Schulke e Zheng 3D}  A. Jaster, J. Mainville, L. Sch\"{u}lke and B.
Zheng, J. Phys. A: Math. Gen. \textbf{32}, 1395 (1999)

\bibitem{Mendes}  J. F. F. Mendes and M. A. Santos, Phys. Rev. \textbf{E57,}
108 (1998)

\bibitem{Drugo e Tania}  T. Tom\'{e} and J. R. Drugowich de Fel\'{i}cio,
Mod. Phys. Lett. \textbf{B12}, 873 (1998)

\bibitem{Grassberger}  P. Grassberger, Physica \textbf{A214,} 547 (1995)

\bibitem{Arcangelis}  L. de Arcangelis, N. Jan J. Phys. \textbf{A19, }L1179%
\textbf{\ }(1986)

\bibitem{Cesar Barreto}  M. Silv\'{e}rio Soares, J. Kamphorst\ Leal da Silva
and F. C. S\'{a} Barreto, Phys. Rev. \textbf{B55}, 1021 (1997)

\bibitem{Ubiraci}  U. P. C. Neves, J. R. Drugowich de Fel\'{i}cio, Physica 
\textbf{A258}, 211 (1998)

\bibitem{Turban2}  L. Turban, Phys. Lett. \textbf{104}, 435 (1984)

\bibitem{Tania}  T. Tom\'{e} and M. J. de Oliveira, Phys. Rev. \textbf{E58,}
4242 (1998)
\end{thebibliography}
\end{document}